\documentclass[preprint,5p,twocolumn]{elsarticle}

\usepackage[utf8]{inputenc}
\usepackage{dcolumn}
\usepackage{amsmath}
\usepackage{amsfonts,amssymb,bm,tensor,braket}
\usepackage{xcolor}
\usepackage{autobreak}
\usepackage{graphicx}
\usepackage{threeparttable,array,booktabs}
\usepackage[referable]{threeparttablex}
\usepackage{makecell, multirow}
\usepackage{tabularx}
\usepackage{setspace}
\usepackage[varg]{txfonts}
\usepackage{caption}
\usepackage[colorlinks=true,allcolors=cyan]{hyperref}
\usepackage[nameinlink,noabbrev]{cleveref}

\bibpunct{\color{cyan}[}{\color{cyan}]}{,}{n}{}{;}

\journal{Results in Physics, published as Results Phys. 26 (2021) 104380, \url{https://doi.org/10.1016/j.rinp.2021.104380} } 

\begin{document}
\hypersetup{allcolors=cyan}

\begin{frontmatter}



\newcommand{\e}{{\mathrm{e}}}
\renewcommand{\i}{{\mathrm{i}}}
\renewcommand{\deg}{^\circ}

\newcommand*{\PKU}{School of Physics and State Key Laboratory of Nuclear Physics and Technology, Peking University, Beijing 100871, China}
\newcommand*{\CHEP}{Center for High Energy Physics, Peking University, Beijing 100871, China}
\newcommand*{\CIC}{Collaborative Innovation Center of Quantum Matter, Beijing, China}

\title{Light Speed Variation with Brane/String-Inspired Space-Time Foam}

\author[a]{Chengyi Li}
\author[a,b,c]{Bo-Qiang Ma\corref{cor1}}

\address[a]{\PKU}
\address[b]{\CHEP}
\address[c]{\CIC}
\cortext[cor1]{Corresponding author \ead{mabq@pku.edu.cn}} 

\begin{abstract}
Recently a series of studies on high energy gamma-ray burst~(GRB) photons suggest a light speed variation with linear energy dependence at the Lorentz violation scale of $3.6 \times 10^{17}~\mathrm{GeV}$, with subluminal propagation of high energy photons in cosmological space. We propose stringy space-time foam as a possible interpretation for this light speed variation. In such a string-inspired scenario, bosonic photon open-string travels \textit{in vacuo} at an infraluminal speed with an energy dependence suppressed by a single power of the string mass scale, due to the foamy structure of space-time at small scales, as described by D-brane objects in string theory. We present a derivation of this deformed propagation speed of the photon field in the infrared (IR) regime. We show that the light speed variation, revealed in the previous studies on GRBs time-delay data, can be well described within such a string approach towards space-time foam. We also derive the value of the effective quantum-gravity mass in this framework, and give a qualitative study on the theory-dependent coefficients. We comment that stringent constraints on Lorentz violation in the photon sector from complementary astrophysical observations can also be explained and understood in the space-time foam context.
\end{abstract}


\end{frontmatter}

\section{Introduction}
\label{Section1}

Constant light speed \textit{in vacuo} is one of the basic postulations in Einstein's theory of special relativity. However, it is speculated that quantum effects of gravity may bring a tiny correction to the speed of light due to Lorentz invariance violation~(LIV) at the Planck scale $M_{\mathrm{Pl}}=\sqrt{hc/2\pi G}\approx 1.22\times 10^{19}~\mathrm{GeV}/c^{2}$~\cite{AmelinoCamelia:2008qg}. Since the light speed variation is extremely small, it is hard to detect such Planck-suppressed LIV modification from most terrestrial experiments. Amelino-Camelia \textit{et al}.~\cite{AmelinoCamelia:1996pj,AmelinoCamelia:1997gz} first proposed that distant celestial sources of high-energy photon emissions, \textit{e.g.}, $\gamma$-ray bursters~(GRBs), can play a crucial role in probing such LIV effects.

Since the launching and the operation of the high-precision \textit{Fermi} $\gamma$-ray space telescope~(FGST), \textit{Fermi}-LAT Collaboration~\cite{Abdo:2009zza,Abdo:2009aa} has reported that with regard to emissions arriving from a GRB, high-energy photons have a tendency to arrive relatively later than low-energy ones. Though such arrival-time retardation of photons may originate from source effects~(non-simultaneous emission at the source) that we know very little, the collaboration~\cite{Abdo:2009aa} also examined the possibilities of vacuum dispersions caused by LIV. Thanks to a number of GRB samples cumulatively observed by the \textit{Fermi} telescope, collective analyses over the whole collection of the photon data can be performed. Inspired by statistically robust methods developed by Ellis \textit{et al}.~\cite{Ellis:1999sd,Ellis:2002in,Ellis:2005wr}, a series of very recent studies on GRB photons have been carried out in Refs.~\cite{Shao:2009bv,Zhang:2014wpb,Xu:2016zxi,Xu:2016zsa,Xu:2018ien,Liu:2018qrg,Amelino-Camelia:2016ohi}. Therein through global analyses of time-of-flight lags between the high-energy photons and the corresponding low-energy signals from different GRBs, 14 \textit{Fermi}-detected very-high-energy events are found to be all compatible with the same LIV energy scale of magnitude of $10^{17}~\mathrm{GeV}$, with 9 of these 14 events lining up very nicely to form a mainline. Such regularity indicates a linearly energy-dependent velocity of light \textit{in vacuo} of the form $c_{g}(\mathcal{E})=1-\mathcal{E}/E_{\mathrm{LIV}}$,\footnote{For brevity, natural units where $\hslash,c\equiv 1$ are used hereafter.} with the first-order Lorentz-violation scale determined as $E_{\mathrm{LIV}}\simeq 3.60\times 10^{17}~\mathrm{GeV}$~\cite{Xu:2016zxi,Xu:2016zsa,Xu:2018ien}. Such finding is certainly impressive as characterized in Refs.~\cite{Xu:2016zsa,Xu:2018ien,Liu:2018qrg,Amelino-Camelia:2016ohi}, it is thus interesting to explore a possible explanation for such finding with a Lorentz-violating framework.

In this paper, we interpret this previously suggested light speed variation as a consequence of stringy interactions of energetic photons from GRBs with the D-particle background incorporated in a Liouville-string scenario of space-time foam~\cite{Ellis:1992eh,Ellis:zh0,Ellis:1999rz,Ellis:zh1,Ellis:2004ay,Ellis:zh2,Ellis:zh3,Li:2009tt,Ellis:2009vq,Li:2021gah}. The basic idea is that the photon, viewed as bosonic open-string state, scatters off the D0-brane space-time defect to form an oscillating intermediate string which subsequently decays to emit the outgoing photon wave with a time delay of $\mathcal{O}\left(1/M_{\mathrm{Pl}}\right)$. The velocity of light in the space-time vacuum is therefore deformed at infrared (IR) relative to the string scale. We indicate that this brane/string-inspired model can serve as a consistent framework to explain the light speed variation as observed from GRBs, in agreement with other astrophysical results to date.

Our paper is organized as follows. In the next section, we briefly review the D-brane realization of space-time foam and its effects on photon propagation. In Sec.~\ref{Section3}, we give a derivation of the deformed speed of light in this string-inspired model. In Sec.~\ref{Section4} we connect observations with the theory to provide an explanation for the finding of the speed-of-light variation as observed from analyses of \textit{Fermi}-GRB data. We then show that some complementary results, including those severe limits from birefringence, photon decay, \textit{etc.}, can also be explained in this framework. In the last section we conclude with some comments on our proposal.

\section{A stringy model for space-time foam}
\label{Section2}

In this section we quickly review the D-brane model of space-time foam and present its physical consequences. The intuitive picture of \textit{space-time foam} goes all the way back to Wheeler~\cite{Wheeler:1998gb}, who noted that there would be highly curved quantum fluctuations with $\Delta E\sim E_{\mathrm{Pl}}$ in the space-time continuum, at short time and small spatial scales $\Delta t\sim 1/E_{\mathrm{Pl}}$, $\Delta x\sim\ell_{\mathrm{Pl}}$, where quantum-gravity effects become relevant.

Within a Liouville non-critical string approach towards quantum gravity (QG)~\cite{Ellis:1992eh,Ellis:zh0,Ellis:1999rz,Ellis:zh1}, a stringy analogue of space-time foam model was developed in Refs.~\cite{Ellis:2004ay,Ellis:zh2}. In this string-inspired framework, the observable~(3+1)-dimensional Universe is described as a compactified D(irichlet)3-brane~\cite{Polchinski:1995mt}, roaming in a higher-dimensional bulk space-time punctured by point-like D0-brane~(``D-particle'') defects in the context of Type-IA string theory~(a T dual of Type-I superstrings~\cite{Schwarz:1999xj}). The standard-model matter particles, propagating through the brane world, are viewed as~(open) string excitations with their ends attached on the D3-brane. When the D3-brane Universe moves in the bulk space, the D-particles traverse it and interact with the string matter anchored on the D3-brane in a topologically non-trivial manner. For an observer living on the brane world, the D-particles ``flash'' on and off, and appear as quantum space-time foam, termed as ``D-foam''~\cite{Ellis:2009vq}.

Due to the presence of D-particle foam defects near the brane world, the dispersion relations for certain string matters would receive ultraviolet (UV) corrections. For massless gauge excitations, like the photon, according to the string formation formalism~\cite{Ellis:zh3} in this framework, each non-trivial interaction between an open photon string and a D-particle excites an oscillating intermediate string state which subsequently decays and re-emits the outgoing photon wave, resulting in an infinitesimal~(causal) time delay for the incident photon propagating through the D-foam. Through modeling the photon-D-particle interactions as stretched intermediate strings or performing straightforward amplitude calculation, one finds that the~(retarded) outgoing waves are emitted with time delays of the order
\begin{equation}\label{eq:2.1}
\Delta t\sim \ell_{s}^{2}\mathcal{E},
\end{equation}
where the string length $\ell_{s}$ is associated with the string mass scale $M_{s}$ through the relation $\ell_{s}=1/M_{s}$, and $\mathcal{E}$ denotes the incident photon energy. The overall time delay consists of each retardation resulting from individual scattering of the photon with a D-particle defect encountered in the foam. Therefore the time delay depends in general on the density $n_{D}$ of foam particles in the bulk space. By taking the linear density of defects $n_{D}/\ell_{s}$ into account, the total delay experienced by a photon can be estimated as
\begin{equation}\label{eq:2.2}
\Delta t_\mathrm{tot}\propto \ell_{s}^{2}{\cal E}\frac{n_{D}}{\ell_{s}}=\frac{\mathcal{E}}{M_{s}}n_{D}.
\end{equation}
In addition to this leading-order delay~\labelcref{eq:2.2}, the collision of the photon with a bulk D-particle would cause the latter to recoil at a velocity of $\mathbf{u}$, which modifies the retarded re-emergence time of the outgoing photon wave to become~\cite{Ellis:zh3}
\begin{equation}\label{eq:2.3}
\Delta t\sim\frac{\ell_{s}^{2}{\cal E}}{1-\lvert{\mathbf{u}}\rvert^{2}},
\end{equation}
which reduces to Eq.~\labelcref{eq:2.1} in the case of $u_{i}\ll1$. In fact the recoil of the D-particle defect would distort the neighboring space-time~\cite{Ellis:zh0,Ellis:zh1}, by inducing an off-diagonal metric element related to the recoil velocity  $u_{i}$ of the D-defect. The resulting space-time fluctuations lead to an \textit{effective} Finsler-like target-space metric as
\begin{equation}\label{eq:2.4}
\eta_{\alpha\beta}^{\mathrm{eff}}=\eta_{\alpha\beta}+h_{\alpha\beta},\ \ h_{0i}=u_{i}.
\end{equation}
Such a distortion of space-time near the recoiling D-particle then affects the~(local) canonical relationship between energy and momentum of the scattered photon through $k^{\alpha}k^{\beta}\eta_{\alpha\beta}^{\mathrm{eff}}=-\mathcal{E}^{2}+2\mathcal{E}k^{i}u_{i}+k^{i}k_{i}=0$, which gives rise to the anomalous dispersion of photons \textit{in vacuo}~\cite{Ellis:2004ay}
\begin{equation}\label{eq:2.5}
\mathcal{E}(\mathbf{k})=\mathbf{k}\cdot\mathbf{u}+\lvert\mathbf{k}\rvert\left(1+\left(\mathbf{u}\cdot\frac{\mathbf{k}}{\lvert\mathbf{k}\rvert}\right)^{2}\right)^{1/2}.
\end{equation}
An important comment is that the delay~\labelcref{eq:2.2} varies linearly with the photon energy and presents no helicity dependence, which stems from the CPT-even feature of the theory.

\section{Velocity dispersion of light in the D-particle quantum-gravitational foam medium}
\label{Section3}

In order to connect the experimental finding with the above theory, in this section we first present a derivation of the modified form of the speed of light within this D-brane QG framework. Since we work in units of $\hslash=1$, the photon dispersion~\labelcref{eq:2.5} can be rewritten, in terms of the frequency $\omega$, as
\begin{equation}\label{eq:3.1}
\omega=\mathbf{k}\cdot\mathbf{u}+k\left(1+\left(\mathbf{u}\cdot \hat{e}_{\mathbf{k}}\right)^{2}\right)^{1/2},
\end{equation}
where $k\equiv\lvert\mathbf{k}\rvert$ is the magnitude of the wave vector, and $\hat{e}_{\mathbf{k}}\equiv\mathbf{k}/k$ denotes the unit vector oriented along the direction of the photon momentum $\mathbf{k}$. In the special case of no D-particle recoil $u_{i}\rightarrow 0$, terms involving the recoil velocity $\mathbf{u}$ in Eq.~\labelcref{eq:3.1} vanish, and the standard relativistic relation for photons $\omega=\lvert\mathbf{k}\rvert$ is recovered. By squaring both sides of Eq.~\labelcref{eq:3.1}, we have
\begin{align}\label{eq:3.2}
\begin{autobreak}
\omega^{2}=\left(\mathbf{k}\cdot\mathbf{u}\right)^{2}
+\left(k\left(1+\left(\mathbf{u}\cdot \hat{e}_{\mathbf{k}}\right)^{2}\right)^{1/2}\right)^{2}
+2k\left(\mathbf{k}\cdot\mathbf{u}\right)\left(1+\left(\mathbf{u}\cdot \hat{e}_{\mathbf{k}}\right)^{2}\right)^{1/2},
\end{autobreak}
\end{align}
where the recoil velocity of the D-particle defect is $u_{i}=M_{s}^{-1}g_{s}\Delta k_{i}=\Delta k_{i}/M_{D}$~\cite{Mavromatos:1998nz}, with $M_{D}=M_{s}/g_{s}$ being the mass for the D-particle. $\Delta k_{i}$ indicates the momentum transfer from the photon to the D-particle. Keep in mind that the string coupling is weak, {\it i.e.}, $g_{s}<1$ is assumed in the model, and that the string scale $M_{s}$ is expected to be a large mass scale, so these D-defects are in general massive objects with very large $M_{D}\sim M_{\mathrm{Pl}}$ in string theory. Therefore a heavy D-particle would recoil with a non-relativistic velocity $u_{i}\ll 1$ after the collision with a photon open-string, as long as the incident energy $k\ll M_{s}$, which we refer to as the IR limit~(relative to the string scale). In this IR region, one can safely expand the last term in Eq.~\labelcref{eq:3.2} to obtain
\begin{align}\label{eq:3.3}
\begin{autobreak}
\omega^{2}=k^{2}
+2k\left(\mathbf{k}\cdot\mathbf{u}\right)
+2\left(\mathbf{k}\cdot\mathbf{u}\right)^{2}
+\left(\mathbf{k}\cdot\mathbf{u}\right)\left(\mathbf{u}\cdot \hat{e}_{\mathbf{k}}\right)^{2}
-\frac{1}{4}k\left(\mathbf{k}\cdot\mathbf{u}\right)\left(\mathbf{u}\cdot \hat{e}_{\mathbf{k}}\right)^{4}
+\mathcal{O}\left(\lvert\mathbf{u}\rvert^{7}\right).
\end{autobreak}
\end{align}
Due to stochastic captures of a photon by D-defects, one should average the above relation~\labelcref{eq:3.3} over the D-particles in the foam, and we adopt the notation
$\langle\!\langle\cdot\!\cdot\!\cdot\rangle\!\rangle_{D}$ to denote this type of averaged values. An anisotropic background field that explicitly breaks Lorentz symmetry then emerges from the average recoil velocity of D-particles $\langle\!\langle u_{i}\rangle\!\rangle_{D}\neq0$~(``anisotropic foam''), whose direction is oriented along that of the incident photon~\cite{Ellis:2000sf}. Note that string theory leads to unambiguous~(minus) sign for $\langle\!\langle u_{i}\rangle\!\rangle_{D}$ due to the fact that there should be no superluminal excitations in a string spectrum. Hence, $\langle\!\langle\mathbf{k}\cdot\mathbf{u}\rangle\!\rangle_{D}=-\langle\!\langle k\lvert\mathbf{u}\rvert\rangle\!\rangle_{D}<0$. Denote the momentum transfer $\Delta k$ in $u_{i}$ as a fraction of the incident momentum $k$, $\Delta k_{i}\equiv \lambda k_{i}$, with $\lambda<1$, we can easily write down
\begin{align}\label{eq:3.4}
\begin{autobreak}
\omega^{2}\simeq k^{2}\Biggl(1
-2\langle\!\langle \lambda\rangle\!\rangle_{D}g_{s}\frac{k}{M_{s}}
+2\langle\!\langle \lambda^{2}\rangle\!\rangle_{D}g_{s}^{2}\frac{k^{2}}{M_{s}^{2}}
-\langle\!\langle \lambda^{3}\rangle\!\rangle_{D}g_{s}^{3}\frac{k^{3}}{M_{s}^{3}}
+\frac{1}{4}\langle\!\langle \lambda^{5}\rangle\!\rangle_{D}g_{s}^{5}\frac{k^{5}}{M_{s}^{5}}
+\mathcal{O}\left(g_{s}^{7}k^{7}/M_{s}^{7}\right)\Biggr),
\end{autobreak}
\end{align}
where averaging notations for $k_{i}$ and $\mathcal{E}$ in the above equation is omitted, also hereafter, for simplicity. The higher-order correction terms are reserved for generality in the expansion~\labelcref{eq:3.4}. To leading order, the modified dispersion relation is given by
\begin{equation}\label{eq:3.5}
\omega^{2}\simeq k^{2}\left(1-2\langle\!\langle \lambda\rangle\!\rangle_{D}g_{s}\frac{k}{M_{s}}\right),
\end{equation}
with $0<\langle\!\langle \lambda\rangle\!\rangle_{D}<1$. Before moving on, we argue that on average the momentum ratio $\lambda$ is related to the bulk density of the D-particle. The total time delay~\labelcref{eq:2.2}, mentioned in Sec.~\ref{Section2}, may be thought of as implying an effective quantum-gravitational \textit{phase} refractive index of the vacuum $\eta_{\mathrm{vac}}(\omega)=1+\mathcal{O}\left(n_{D}\omega/M_{D}\right)>1$ for light with frequency $\omega$. By substituting it into $c_{\mathrm{ph}}\equiv1/\eta$,\footnote{Here $c_{\mathrm{ph}}\equiv c/\eta=1/\eta$ with $c=1$ being the conventional speed of light in normal vacuo of Special Relativity (\textit{i.e.}, the propagation velocity of low energy photons). Recall that in Lorentz invariant space-time, the group velocity and the phase velocity of photons in vacuum are the same, which equal to the constant speed of light, $c_{g}=c_{\mathrm{ph}}=1$.} the phase velocity of photons in the QG D-foam medium can be estimated as $c_{\mathrm{ph}}(k)\simeq 1-\left\lvert\mathcal{O}\left(n_{D}k/M_{D}\right)\right\rvert$. On the other hand, from equation~\labelcref{eq:3.5} one can derive the same phase velocity as $c_{\mathrm{ph}}=\omega/\lvert\mathbf{k}\rvert\simeq 1-\langle\!\langle\lambda\rangle\!\rangle_{D}g_{s}k/M_{s}$. This fact yields the expected relation
\begin{equation}\label{eq:3.6}
\langle\!\langle\lambda\rangle\!\rangle_{D}\propto n_{D}.
\end{equation}
Assume that the relation $c_{g}=\partial\omega/\partial k$ still holds in QG, the propagation velocity of light can be straightforwardly deduced, on account of Eq.~\labelcref{eq:3.4}, as follows
\begin{align}\label{eq:3.7}
\begin{autobreak}
c_{g}=\frac{k}{\omega}\Biggl(1
-3\langle\!\langle\lambda\rangle\!\rangle_{D}g_{s}\frac{k}{M_{s}}
+4\langle\!\langle \lambda^{2}\rangle\!\rangle_{D}g_{s}^{2}\frac{k^{2}}{M_{s}^{2}}
-\frac{5}{2}\langle\!\langle \lambda^{3}\rangle\!\rangle_{D}g_{s}^{3}\frac{k^{3}}{M_{s}^{3}}
+\mathcal{O}\left((k/M_{s})^{5}\right)\Biggr),
\end{autobreak}
\end{align}
where $k/\omega$ factor is given by
\begin{align}\label{eq:3.8}
\frac{k}{\omega}\simeq&\ \Biggl(1
-2\langle\!\langle \lambda\rangle\!\rangle_{D}g_{s}\frac{k}{M_{s}}
+2\langle\!\langle \lambda^{2}\rangle\!\rangle_{D}g_{s}^{2}\frac{k^{2}}{M_{s}^{2}}
-\langle\!\langle \lambda^{3}\rangle\!\rangle_{D}g_{s}^{3}\frac{k^{3}}{M_{s}^{3}}\Biggr)^{-1/2}\nonumber\\
=&\ 1+\langle\!\langle \lambda\rangle\!\rangle_{D}g_{s}\frac{k}{M_{s}}+\frac{1}{2}\langle\!\langle \lambda^{2}\rangle\!\rangle_{D}g_{s}^{2}\frac{k^{2}}{M_{s}^{2}}+\mathcal{O}\left((k/M_{s})^{4}\right).
\end{align}
So the group velocity of photons in a D-particle foam background is modified by
\begin{equation}\label{eq:3.9}
c_{g}=1-2\langle\!\langle\lambda\rangle\!\rangle_{D}g_{s}\frac{k}{M_{s}}+\frac{3}{2}\langle\!\langle \lambda^{2}\rangle\!\rangle_{D}g_{s}^{2}\frac{k^{2}}{M_{s}^{2}}+\mathcal{O}\left(k^{4}/M_{s}^{4}\right).
\end{equation}
Note that terms proportional to odd powers of momentum, like the $k^{3}/M_{s}^{3}$ term, in Eq.~\labelcref{eq:3.7}, cancel out in the final expression~\labelcref{eq:3.9} except for the single-power term, which is a striking feature of this stringy approach to QG-dispersed velocity of light, and we shall comment on that from another viewpoint below. With the IR condition~($k\ll M_{s}$) applied, we have
\begin{equation}\label{eq:3.10}
c_{g}\simeq1-2\langle\!\langle\lambda\rangle\!\rangle_{D}g_{s}\frac{k}{M_{s}}.
\end{equation}

We comment at this point that there is another way to have the photon speed obtained. Instead of deriving the group velocity from the squared form of the dispersion relation~\labelcref{eq:3.4}, one can start directly from Eq.~\labelcref{eq:3.1}. Utilizing the relation $\langle\!\langle\mathbf{k}\cdot\mathbf{u}\rangle\!\rangle_{D}\simeq-M_{s}^{-1}g_{s}\langle\!\langle\lambda\rangle\!\rangle_{D} k^{2}$, we have the exact formula of the velocity
\begin{align}\label{eq:3.11}
\begin{autobreak}
c_{g}=1-2\langle\!\langle\lambda\rangle\!\rangle_{D}g_{s}\frac{k}{M_{s}}
+\langle\!\langle~\Biggl(1+\lambda^{2}g_{s}^{2}\frac{k^{2}}{M_{s}^{2}}\Biggr)^{1/2}-1\rangle\!\rangle_{D}
+\langle\!\langle~\Biggl(1+\lambda^{2}g_{s}^{2}\frac{k^{2}}{M_{s}^{2}}\Biggr)^{-1/2}\lambda^{2} g_{s}^{2}\frac{k^{2}}{M_{s}^{2}}\rangle\!\rangle_{D}.
\end{autobreak}
\end{align}
In the IR regime, the latter two terms in the above equation can be expressed as the Taylor series expansion, and one can easily check that this equation then reduces to Eq.~\labelcref{eq:3.9}. We see that the second term in~\labelcref{eq:3.11} just corresponds to the linear-order correction to light speed \labelcref{eq:3.10}, and it is reasonable that odd-power terms $\mathcal{O}\left((k/M_{s})^{2n+1}\right)~(n>0)$ do not exist since the Taylor expansion would only give birth to those terms suppressed by even powers of the string mass $M_{s}$.

We stress again that equations~\labelcref{eq:3.9} as well as~\labelcref{eq:3.10} are valid only if $\langle\!\langle\lvert\mathbf{u}\rvert\rangle\!\rangle_{D}\ll 1$, corresponding to $k\ll M_{s}$. Our new-developed formula~\labelcref{eq:3.11} seemingly does not rely on this IR limit however, so it is likely to work even when $k\sim M_{s}$. But one should notice that Eq.~\labelcref{eq:3.11} together with our starting point~\labelcref{eq:2.5} actually depend on the metric deformation formalism~\labelcref{eq:2.4} which naturally requires the distortion of space-time not to be strong, namely $h_{0i}=u_{i}\ll1$. Hence in the case of relativistic recoil of D-particles $u_{i}\rightarrow 1$ that could be attained when the incident energy $k\sim M_{s}$, Eq.~\labelcref{eq:3.9} and the precise form~\labelcref{eq:3.11} of the velocity are no longer valid. For this reason, in the UV regime one should stick to the string formation formalism. At energies $k\sim M_{s}/g_{s}$, as a matter of fact the recoiling D-particle moves at nearly the relativistic speed of light, $\lvert\mathbf{u}\rvert^{2}=\mathcal{O}\left(1\right)$, as a result the time delay $\Delta t$ goes to infinity (\textit{c.f.}~\labelcref{eq:2.3}), which is a clear manifestation of \textit{the destabilization of the vacuum}. This fact implies that the recoil of the D-particle, when scattered off such ultra-high-energy photons, would induce a very strong space-time distortion, leading to complete \textit{absorption} of such photons by the D-particle defect~\cite{Ellis:2010he}. The defect behaves like a black hole, capturing permanently the incident photons~\cite{Mavromatos:zh}.

Before closing this section, we define, for convenience in the discussions that follow, the~(stringy) quantum-gravity ``effective'' scale for photons in the D-particle foam by
\begin{equation}\label{eq:3.12}
\mathcal{M}_{\mathrm{QG}\gamma}^{(\textrm{D-foam})}=\frac{M_{s}}{2\langle\!\langle\lambda\rangle\!\rangle_{D}g_{s}}\equiv M_{s}/\varsigma_{\gamma,\mathrm{ani}}^{(\textrm{D-foam})},
\end{equation}
where $\varsigma_{\gamma,\mathrm{ani}}^{(\textrm{D-foam})}=2\langle\!\langle\lambda\rangle\!\rangle_{D}g_{s}>0$ denotes a nonzero dimensionless coefficient for the anisotropic D-foam.\footnote{We consider in this paper a \textit{uniform} density of D-particles, \textit{i.e.}, $\varsigma_{\gamma,\mathrm{ani}}^{(\textrm{D-foam})}$ is a redshift-$z$-independent coefficient. In more complicated cases, the density $n_{D}$ could vary with the cosmological epoch~\cite{Ellis:2009vq,Mavromatos:2009mj}, then the effective QG scale could depend also on the redshift: $\widetilde{\mathcal{M}}_{\mathrm{QG}\gamma}^{(\textrm{D-foam})}(z)\sim n_{D}(z)/M_{s}$.} We emphasis here that this relevant QG scale~\labelcref{eq:3.12} cannot be \textit{a priori} determined.  To see this clearly, one should note that, though the underlying string theory, the most promising framework for \textit{a theory of everything}, incorporating the compelling description of gravity quantization, should have no free parameters, the scale $M_{s}$, according to modern understanding in string theory~\cite{Polchinski:1998zh}, is indeed an adjustable constant to be constrained by studies on string phenomenology. Moreover, the coefficient $\varsigma_{\gamma,\mathrm{ani}}^{(\textrm{D-foam})}$ that appears in the QG scale~\labelcref{eq:3.12} relates to the microscopic setup of the model. Specifically, we see that Eq.~\labelcref{eq:3.6} leads to a relation $\varsigma_{\gamma,\mathrm{ani}}^{(\textrm{D-foam})}\sim g_{s}n_{D}$, where the D-particle density $n_{D}$ in the bulk is arbitrary in general, implying that it is essentially a free parameter in the model~\cite{Mavromatos:2009xg}. Hence the scale~\labelcref{eq:3.12} is free to vary, and thus needs to be constrained by experiments. Given the effective QG mass defined above, the deformed velocity of light \labelcref{eq:3.10} can be expressed as the following simple form
\begin{equation}\label{eq:3.13}
c_{g}\simeq 1-\omega/\mathcal{M}_{\mathrm{QG}\gamma}^{(\textrm{D-foam})},
\end{equation}
where, to leading order, $k\simeq\omega$. In the next section this QG-modified speed of light~\labelcref{eq:3.13} can be used to explain the experimental finding in Refs.~\cite{Xu:2016zxi,Xu:2016zsa,Xu:2018ien} from the GRB data.

\section{An explanation for the light speed variation in GRBs and further discussions}
\label{Section4}

Since the launching and the operation of the high-precision FGST, the \textit{Fermi}-LAT Collaboration has reported that high energy photons arrive at the Earth later relative to low energy ones~\cite{Abdo:2009zza,Abdo:2009aa}. Though such time-of-flight lags may be attributed to source-intrinsic properties of GRBs, there are still possibilities that arrival time delays originate from the speed dispersions of photons duration the propagation, arising from LIV. In the last decade, a number of energetic GRB samples have been cumulatively obtained by the \textit{Fermi} telescope, making it possible to perform global analyses on a range of GRBs to disentangle LIV-induced lag from \textit{a priori} unknown source-intrinsic lag and further reveal fascinating regularities which can serve as the supports for the light speed variation.

Intriguing statistical analyses of photon time-of-flight over the whole collection of \textit{Fermi}-detected GRBs have been carried out in a series of very recent studies~\cite{Shao:2009bv,Zhang:2014wpb,Xu:2016zxi,Xu:2016zsa,Xu:2018ien,Liu:2018qrg,Amelino-Camelia:2016ohi}. In Refs.~\cite{Xu:2016zxi,Xu:2016zsa,Xu:2018ien}, utilizing a general Lorentz-violation modified dispersion relation with vanishing rest mass
\begin{equation}\label{eq:4.1}
\mathcal{E}^{2}\simeq p^{2}-s_{n}\mathcal{E}^{2}\left(\frac{p}{E_{\mathrm{LIV},n}}\right)^n,
\end{equation}
the authors obtain the deformed photon propagation velocity with higher-order terms dropped as
\begin{equation}\label{eq:4.2}
v_{g}(\mathcal{E})=\frac{\partial\mathcal{E}}{\partial p}\simeq 1-s_{n}\frac{n+1}{2}\left(\frac{\mathcal{E}}{E_{\mathrm{LIV},n}}\right)^n,
\end{equation}
where $n=1$ or $2$ corresponds to linear or quadratic energy dependence respectively, $s_{n}=\pm1$ is a sign factor of LIV correction, and $E_{\mathrm{LIV},n}$ is the $n$th-order Lorentz-violation scale. Using the arrival-time differences of multi-GeV photon signals from a number of $\textit{Fermi}$-observations, a regularity fitting well with all 14 high-energy events is observed by the authors, indicating a first-order Lorentz-violation scale~(\textit{i.e.}, $n=1$) at\footnote{Compared to the quadratic ($n=2$) LIV case, as illustrated in Ref.~\cite{Xu:2016zxi}, the linear ($n=1$)  case is more favored by the photon time-delay data. In the case $n=1$ the redundant subscripts $n$ in the above formulae are hereafter omitted.}
\begin{equation}\label{eq:4.3}
E_{\mathrm{LIV}}=\left(3.60\pm0.26\right)\times10^{17}~\mathrm{GeV},
\end{equation}
which is close to the Planck scale $E_{\mathrm{Pl}}\simeq1.22\times10^{19}~\mathrm{GeV}$. A positive sign factor $s=+1$ is suggested through fitting the data, implying that high-energy photons travel across cosmos slower than their low-energy counterparts. More striking regularity is that 9 out of these 14 photons falls very nicely on a same mainline, giving a strong indication of a light speed variation suppressed by a single power of the Planck scale.\footnote{Statistical significance of such finding has been studied in Refs.~\cite{Xu:2018ien,Liu:2018qrg,Amelino-Camelia:2016ohi}, \textit{e.g.}, one finds that this light speed variation is flavored at $3$-$5\sigma$ confidence level (C.L.) in Ref.~\cite{Xu:2018ien}, where we refer the interested reader for further details.} It is worth noting that such a LIV scale~\labelcref{eq:4.3} is consistent with various constraints from high-energy $\gamma$-ray observations of pulsars~\cite{Kaaret:1999ve,Otte:2012tw,Zitzer:2013gka,Chretien:2015thesis}, active galactic nuclei (AGNs)~\cite{Biller:1998hg,Albert:2007qk,Martinez:2008ki,Romoli:2017zuc,Abdalla:2019krx} as well as GRBs~\cite{Ellis:1999sd,Ellis:2002in,Ellis:2005wr,Boggs:2003kxa,Bolmont:2006kk,Chang:2015qpa,Bernardini:2017tzu,Pan:2020zbl,Du:2020uev} and it is also compatible with the strongest robust limit to date~\cite{Ellis:2018lca} from a recent study on 8 \textit{Fermi}-LAT GRBs with bright emissions in multi-$\mathrm{GeV}$ energies~\cite{Ackermann:2012kna}. We need to mention that there can be more stringent limits on this characteristic scale of LIV from individual analyses of flaring PKS~2155-304~\cite{Aharonian:2008kz,HESS:2011aa},\footnote{In Ref.~\cite{Aharonian:2008kz}, time delays between two light curves of different energy bands of PKS~2155-304 were used to give a lower bound of $E_{\mathrm{LIV}}>7.2\times 10^{17}~\mathrm{GeV}$. However a latest work~\cite{Li:2020uef} points out a novel way to understand the delays which can be considered as a signal for the light speed variation with $E_{\mathrm{LIV}}\simeq 3.6\times 10^{17}~\mathrm{GeV}$, \textit{i.e.}, Eq.~\labelcref{eq:4.3}. We refer the reader to~\cite{Li:2020uef} for details.} short GRB~090510~\cite{Abdo:2009aa,Vasileiou:2013vra,Vasileiou:2015wja} or striking TeV event of GRB~190114C lately found by MAGIC~\cite{Acciari:zh,Acciari:2020kpi}, which tend to place lower bounds $E_{\mathrm{LIV}}\gtrsim\left(0.1-10\right)\times E_{\mathrm{Pl}}$, stronger than that in Eq.~\labelcref{eq:4.3}. Therefore more observations are needed to further verify the speed of light modification proposed in Refs.~\cite{Xu:2016zxi,Xu:2016zsa,Xu:2018ien}.

We discuss now the implications of the above phenomenological results from a theoretical perspective. In Sec.~\ref{Section3} we saw that the dispersion relation and the speed of light of the photon field are modified due to the interaction of photon open-string with the D-particle string foam. We indicate in the following that this fact suggests a natural way to explain the speed variation of light mentioned above.

Note that one of the most interesting features revealed in Refs.~\cite{Xu:2016zxi,Xu:2016zsa,Xu:2018ien} is that multi-GeV events from different samples of GRBs line up surprisingly to form a mainline, indicating a \textit{linear energy dependence} in the speed of light, $\lvert v_{g}-1\rvert\sim\mathcal{O}\left(\mathcal{E}/E_{\mathrm{LIV}}\right)$, rather than a quadratic~(or higher-order) dependence of the photon energy, with $E_{\mathrm{LIV}}$ characterizing such a linear suppression of LIV determined to be a few $10^{17}~\mathrm{GeV}$, \textit{i.e.}, Eq.~\labelcref{eq:4.3}, 2-order less than the Planck energy scale. If we consider this result in terms of the phase velocity $v_{\mathrm{ph}}\equiv 1/\eta=\mathcal{E}/p$, taking into account the modified dispersion relation~\labelcref{eq:4.1}, with
\begin{equation}\label{eq:4.4}
\eta=1+\frac{1}{2}s_{n}\left(\frac{\mathcal{E}}{E_{\mathrm{LIV},n}}\right)^{n}.
\end{equation}
the light speed variation with $E_{\mathrm{LIV}}\simeq 3.60\times 10^{17}~\mathrm{GeV}$ just corresponds to the linear case, \textit{i.e.}, $n=1$ in Eq.~\labelcref{eq:4.4}, which reduces to $\eta=1+\frac{1}{2}\frac{\mathcal{E}}{E_{\mathrm{LIV}}}$. Thus the finding in Refs.~\cite{Xu:2016zxi,Xu:2016zsa,Xu:2018ien} implies that current data favor such a vacuum refractive index that varies linearly with the energy of the photon. While according to the D-brane-inspired space-time foam~(\textit{c.f.}, Sec.~\ref{Section2}), the non-relativistic recoil of the foam defect when interacts with an energetic photon propagating over our brane-world Universe can lead to a deformed background space-time metric ``felt'' by the photon. As a result, the velocity of light receives a QG-induced UV correction which grows linearly on the photon energy, $c_{g}\sim 1-\mathcal{E}/\mathcal{M}_{\mathrm{QG}\gamma}^{(\textrm{D-foam})}$, \textit{i.e.}, Eq.~\labelcref{eq:3.13}, with $\mathcal{M}_{\mathrm{QG}\gamma}^{(\textrm{D-foam})}$ approaching the Planck mass. Due to the presence of the D-foam, the space-time vacuum appears no longer transparent to photons, but behaves essentially like a dispersive medium with non-trivial refractive indices $\eta_{\mathrm{vac}}(\mathcal{E})=1+\mathcal{O}\left(\mathcal{E}/\mathcal{M}_{\mathrm{QG}\gamma}^{(\textrm{D-foam})}\right)$ for photon propagation. Obviously, these features of the string foam model is exactly consistent with the linear form energy-dependence of light speed observed by the authors of Refs.~\cite{Xu:2016zxi,Xu:2016zsa,Xu:2018ien}.

Furthermore, we saw that the result in these previous works leads also to a positive sign factor~(\textit {i.e.}, $s=+1$) for the linear LIV correction, indicating the fact that photons with higher energies propagate more slowly than lower ones, corresponding to a \textit{subluminal photon propagation}.\footnote{It might be argued that the non-observation of superluminal photon propagation in Refs.~\cite{Xu:2016zxi,Xu:2016zsa,Xu:2018ien} may be attributed to $\gamma$-decays allowed by a modified superluminal dispersion effect. However such superluminal vacuum dispersion has been severely constrained by the absence of $\gamma$-decay~\cite{Liberati:2009pf,Shao:2010wk}, which, as demonstrated below, can serve as an evidence to support the D-particle foam interpretation of light speed variation presented in this paper.} From the viewpoint of the phase velocity, this result implies $\eta>1$ in Eq.~\labelcref{eq:4.4}. Within the D-brane foam context, the capture and re-emission processes of the photon by D-particle defects induce a causal time delay~\labelcref{eq:2.2}, resulting in a deceleration of high-energy photons by the D-particle foam. This effect manifests itself through a subluminal velocity $c_{g}(\mathcal{E})\simeq 1-\varsigma_{\gamma,\mathrm{ani}}^{(\textrm{D-foam})}\mathcal{E}/M_{s}$ with the quantity $\varsigma_{\gamma,\mathrm{ani}}^{(\textrm{D-foam})}/M_{s}\sim\langle\!\langle\lambda\rangle\!\rangle_{D}/M_{D}$ (denoted by $\mathrm{LIV}_{\textrm{D-foam}}$) being positive by construction~(\textit{c.f.}, Sec.~\ref{Section3}). Meanwhile, such a time delay \labelcref{eq:2.2} may be thought of as implying an increasing refractive index of light, $\eta_{\mathrm{vac}}=1+\mathcal{O}\left(\ell_{s}\mathcal{E}\right)>1$, with photon energy. As a result, the more energetic one photon is, the slower it travels through space. It becomes clear that such prediction of the space-time foam model is consistent with the subluminal feature of photons, as exposed in Refs.~\cite{Xu:2016zxi,Xu:2016zsa,Xu:2018ien} from GRB observations.

With the established consistency between the propagation velocity from the generalized LIV-modified dispersion relation in Eq.~\labelcref{eq:4.2} with the pair~$(n, s_{n})$ setting to~$(1, +1)$ and the velocity from the stringy space-time foam scenario in Eqs.~\labelcref{eq:3.10} and~\labelcref{eq:3.13}, we can relate $\mathcal{M}_{\mathrm{QG}\gamma}^{(\textrm{D-foam})}$ to the Lorentz-violation scale $E_{\mathrm{LIV}}$. On account of Eq.~\labelcref{eq:4.3}, we arrive at a relation
\begin{equation}\label{eq:4.5}
\mathcal{M}_{\mathrm{QG}\gamma}^{(\textrm{D-foam})}\simeq E_{\mathrm{LIV}}\simeq 3.60\times 10^{17}~\mathrm{GeV},
\end{equation}
or, equivalently we have
\begin{equation}\label{eq:4.6}
\frac{\langle\!\langle\lambda\rangle\!\rangle_{D}g_{s}}{M_{s}}\simeq\frac{1}{2E_{\mathrm{LIV}}}\simeq 1.39\times 10^{-18}~\mathrm{GeV}^{-1}.
\end{equation}
The result~\labelcref{eq:4.5} leads to a claim of an effective quantum-gravity scale $\sim 10^{17}~\mathrm{GeV}$ for the string/D-particle foam model. This is very close to the expectation that this scale is taken to be very large, typically an order of magnitude or so below the Planck scale, $\mathcal{M}_{\mathrm{QG}\gamma}^{(\textrm{D-foam})}\lesssim\mathcal{M}_{\mathrm{Pl}}=\mathcal{O}\left(10^{18}\right)~\mathrm{GeV}$, where $\mathcal{M}_{\mathrm{Pl}}=\sqrt{1/8\pi G}\approx 2\times 10^{18}~\mathrm{GeV}$ is the reduced Planck mass. If we take this result as the lower bound on the scale of stringy quantum gravity in the space-time foam scenario, then our result~\labelcref{eq:4.5} is very similar to that obtained recently in Ref.~\cite{Ellis:2018lca}. Given that $\langle\!\langle\lambda\rangle\!\rangle_{D}\lesssim\mathcal{O}\left(1\right)$, we can further deduce the constraint
\begin{equation}\label{eq:4.7}
\frac{M_{s}}{g_{s}}<7.20\times 10^{17}~\mathrm{GeV},
\end{equation}
which is compatible with the natural assumption~\cite{Mavromatos:2009mj} for the value $M_{s}/g_{s}\sim 10^{19}~\mathrm{GeV}$ of the mass of the foam defect. Meanwhile, we find that our result leads to a rough estimate about the scale $M_{s}\sim 3.60\times 10^{17}~\mathrm{GeV}$, which gives a string mass scale nearly $0.15\times \mathcal{M}_{\mathrm{Pl}}$ if $\varsigma_{\gamma,\mathrm{ani}}^{(\textrm{D-foam})}$ is of order $1$. While in general, taking Eq.~\labelcref{eq:4.6} into account, we get
\begin{equation}\label{eq:4.8}
M_{s}\simeq 7.20\times 10^{17}\langle\!\langle\lambda\rangle\!\rangle_{D}g_{s}~\mathrm{GeV}.
\end{equation}
If $\langle\!\langle\lambda\rangle\!\rangle_{D}g_{s}\sim\mathcal{O}\left(1\right)$ and then an intriguing estimation of $M_{s}$ is at the scale of $\mathrm{a~few}\times 10^{17}~\mathrm{GeV}$, which is consistent with the expectation in modern string theory that $M_{s}\neq M_{\mathrm{Pl}}$ in general. While usually it is assumed in the string foam model that the combination $\langle\!\langle\lambda\rangle\!\rangle_{D}g_{s}$ is (much) lower than unity, then the string scale $M_{s}$ is roughly not higher than $10^{16}~\mathrm{GeV}$ and it can be as low as a few $\mathrm{TeV}$, which could be attained if $\langle\!\langle\lambda\rangle\!\rangle_{D}\ll 1$ or $g_{s}\ll 1$. In this case, the result may be compatible with several schemes of low-scale string~\cite{Antoniadis:1999rm,Antoniadis:2000vd} where the mass scale $M_{s}$ can be decreased arbitrarily low, though it cannot be lower than $\sim 10^{4}~\mathrm{GeV}$ since if this were the case we should have already observed fundamental strings in the experiments at the Large Hadron Collider (LHC). In addition, we also obtain analogous estimations of the string length scale $\ell_{s}$ and the universal Regge slope $\alpha^{\prime}$ as
\begin{align}\label{eq:4.8}
\ell_{s}\simeq&\ 1.39\times 10^{-18}\frac{1}{\langle\!\langle\lambda\rangle\!\rangle_{D}g_{s}}~\mathrm{GeV}^{-1},\\
\alpha^{\prime}\simeq&\ 1.93\times 10^{-36}\frac{1}{\langle\!\langle\lambda\rangle\!\rangle_{D}^{2}g_{s}^{2}}~\mathrm{GeV}^{-2}.
\end{align}
For fundamental string length $\ell_{s}$, our estimate~\labelcref{eq:4.8} further implies $\ell_{s}\simeq 1.69\times 10^{-19}\frac{1}{\langle\!\langle\lambda\rangle\!\rangle_{D}g_{s}}~\mathrm{GeV}^{-1}\left(\frac{\ell_{\mathrm{Pl}}}{10^{-28}~\mathrm{eV}^{-1}}\right)\approx 1.7\frac{\ell_{\mathrm{Pl}}}{\langle\!\langle\lambda\rangle\!\rangle_{D}g_{s}}\sim\ell_{\mathrm{Pl}}/n_{D}g_{s}$, with the Planck length $\ell_{\mathrm{Pl}}\simeq 1.6\times 10^{-35}~\mathrm{m}$.

From the above discussion we see that the light speed variation proposed previously from GRB photons~\cite{Xu:2016zxi,Xu:2016zsa,Xu:2018ien} can be described in a D-brane/string-inspired model of space-time foam with $\mathcal{M}_{\mathrm{QG}\gamma}^{(\textrm{D-foam})}\simeq 10^{17}~\mathrm{GeV}$. While there are still several classes of complementary (astrophysical) constraints on LIV in the photon sector, which need to be considered in our attempt to interpret the finding in Refs.~\cite{Xu:2016zxi,Xu:2016zsa,Xu:2018ien}. These constraints could come from birefringence effects as well as photon decay processes. We indicate below that these results can also be understood in the framework of D-particle space-time foam, as shown in Ref.~\cite{Li:2021gah}.

\begin{table}
\centering
\caption{Some constraints on LIV-induced vacuum birefringence from polarimetric observations of astrophysical sources. Lower bounds on LIV scale for birefringence propagation $E_{\mathrm{LIV}}^{(\mathrm{bire})}\gtrsim{\bm\varepsilon}=\left(4\ell_{\mathrm{Pl}}\lvert\chi\rvert_{\mathrm{upper}}\right)^{-1}$ are calculated.}
\label{tab:4.1}
\begin{threeparttable}
\def\arraystretch{1.3}
\begin{tabular}{l l l l}
\toprule\midrule
Source & Distance & $\lvert\chi\rvert_{\mathrm{upper}}$\tnotex{tn:a} & ${\bm\varepsilon}~(\mathrm{GeV})$\\
\midrule
3C 256 & $z\simeq1.82$ & $5\times10^{-5}$~\cite{Gleiser:2001rm} & $6.1\times10^{22}$\\
GRB~020813 & $z\simeq1.26$ & $3\times10^{-7}$~\cite{Fan:2007zb} & $1.0\times10^{25}$\\
Crab pulsar & $\sim1.9~\mathrm{kpc}$ & $2.3\times10^{-10}$~\cite{Maccione:2008tq} & $1.3\times10^{28}$\\
Cygnus X-1 & $\sim2.2~\mathrm{kpc}$ & $8.7\times10^{-12}$~\cite{Shao:2011uc} & $3.5\times10^{29}$\\
GRB~041219A & $z\sim0.31$ & $5.5\times10^{-15}$~\cite{Laurent:2011he} & $5.5\times10^{32}$\\
GRB~110721A & $z\simeq0.38$ & $2.6\times10^{-16}$~\cite{Wei:2019nhm} & $1.2\times10^{34}$\\
GRB~160509A & $z\simeq1.17$ & $4\times10^{-17}$~\cite{Wei:2019nhm}\tnotex{tn:b} & $7.6\times10^{34}$\\
GRB~061122 & $z\simeq1.33$ & $2.5\times10^{-17}$~\cite{Lin:2016xwj} & $1.2\times10^{35}$\\
\bottomrule
\end{tabular}
\begin{tablenotes}
\footnotesize
\item [a] \label{tn:a} The numerical factors between our notation from those of quoted references are accounted.
\item [b] \label{tn:b} In Ref.~\cite{Wei:2019nhm}, 12 latest polarized GRBs were assembled to place strict limits on $\chi$ in the order of $\mathcal{O}\left(10^{-15}-10^{-17}\right)$. We refer the reader to~\cite{Wei:2019nhm} for more details on these impressive results.
\end{tablenotes}
\end{threeparttable}
\end{table}

We focus first on strong constraints based on vacuum birefringence. Birefringence is predicted in certain models of QG, with both Lorentz invariance and reflexion symmetry (parity/CPT) violation, where the velocity variation of two polarization modes (denoted by $h_{\pm}$) can be parametrized by
\begin{equation}\label{eq:4.9}
\delta v_{g}(\mathcal{E},h_{\pm})=\mp4\chi\mathcal{E}/M_{\mathrm{Pl}}.
\end{equation}
In Eq.~\labelcref{eq:4.9} $\chi$ is a parameter following from the underlying theory, which characterizes the helicity dependence of photon velocity and governs CPT-odd LIV effects. Limits on the parameter $\chi$ can be interpreted in terms of an analogous LIV energy scale for birefringence propagation
\begin{equation}\label{eq:4.10}
E_{\mathrm{LIV}}^{(\mathrm{bire})}=\left\lvert 4\chi\ell_{\mathrm{Pl}}\right\rvert^{-1}.
\end{equation}
Strong constraints on $\chi$ (or $E_{\mathrm{LIV}}^{(\mathrm{bire})}$) have been set in astro-particle physics by wide studies on polarization measurements from various sources. In \Cref{tab:4.1}, we list some results obtained in the last two decades. For example, the observation of polarized lights from the Crab Nebula constrains $\mathcal{O}\left({\mathcal{E}}/M_{\mathrm{Pl}}\right)$ LIV in Eq.~\labelcref{eq:4.9} to the level $\lvert\chi\rvert<2\times 10^{-10}$ at $95\%$ C.L.~\cite{Maccione:2008tq}, which corresponds to $E_{\mathrm{LIV}}^{(\mathrm{bire})}\gtrsim 1\times 10^{28}~\mathrm{GeV}$. The most stringent constraint comes from GRB 061122 and the birefringence parameter $\chi$ is determined to be $\lvert\chi\rvert\lesssim 10^{-17}$~\cite{Lin:2016xwj}, tightening by about seven orders of magnitude the Crab Nebula constraint. Hence one finds from the table that polarimetric observations of light from remote astronomical sources put very strong limits on a modified \textit{birefringent} speed of light~\labelcref{eq:4.9} and almost rule out LIV-motivated birefringence effects. While in the stringy space-time foam scenario, the coupling of the photons to the D-foam medium is independent of photon polarization. Specifically, we saw in Eqs.~\labelcref{eq:2.1},~\labelcref{eq:2.2} and~\labelcref{eq:3.5}~(or Eq.~\labelcref{eq:3.10}) that string calculations \textit{do not} yield a dependence of the speed of light on its polarization mode, thereby the coefficients~($\varsigma_{\gamma,\mathrm{ani}}^{(\textrm{D-foam})}$) for left- and right-handed circularly polarized states are necessarily the same, $\varsigma_{\gamma,\mathrm{ani},h_{+}}^{(\textrm{D-foam})}=\varsigma_{\gamma,\mathrm{ani},h_{-}}^{(\textrm{D-foam})}$. Photons with opposite ``helicities'' but the same frequency have the same amounts of speed variation
\begin{equation}\label{eq:4.11}
\delta c_{g}(\mathcal{E})\simeq -n_{D}g_{s}\mathcal{E}/M_{s},~\text{for states}~~h_{+}, h_{-},
\end{equation}
and as a result, the usual degeneracy of $c_{g}$ amongst polarizations is preserved when Lorentz symmetry is broken. Birefringence is therefore not expected to present in this parity-even theory. This fact leads to a natural way to account for the above-mentioned lower limits for $E_{\mathrm{LIV}}^{(\mathrm{bire})}$. Since the D-particle string foam does not predict vacuum birefringence, the polarimetric observation cannot be used to bound $\mathrm{LIV}_{\textrm{D-foam}}$. Thus, those tight restrictions on $E_{\mathrm{LIV}}^{(\mathrm{bire})}$ calculated in~\Cref{tab:4.1} cannot translate to upper limits for the quantity $n_{D}g_{s}/M_{s}$ in Eq.~\labelcref{eq:4.11}, and in such a way that the above stringent constraints are no longer valid, thereby producing \textit{no} incompatibility with that given in Eqs.~\labelcref{eq:4.5} and~\labelcref{eq:4.6}. On the other hand, polarized light measurements do result in severe limits on linearly suppressed LIV, as seen from~\Cref{tab:4.1} (\textit{i.e.}, for linear models~\labelcref{eq:4.9} the lower bound on $E_{\mathrm{LIV}}^{(\mathrm{bire})}$ exceeds the Planck scale by as much as sixteen orders of magnitude). However, with the stringy foam model, one can easily understand the reason for that. Indeed, if the vacuum birefringence effect does not exist at all, as theoretically predicted, then it would be natural to expect that extremely tight constraints on any effects induced by a helicity-$h_{\pm}$-dependent speed variation~\labelcref{eq:4.9} should be placed by experiments. This is exactly the situation for current observational results. Therefore, it is clear that the D-particle foam model is capable of explaining well the constraints from the absence of detectable birefringence.

Moreover, following that same logic, we can interpret the results from another  LIV phenomenon, namely the photon decay, $\gamma\rightarrow e^{+}e^{-}$, allowed by superluminal LIV in Eq.~\labelcref{eq:4.9}. Since photon decay in flight from the source would lead to a hard cutoff in astrophysical $\gamma$-ray spectra~\cite{Liberati:2001cr}, a lower bound for $E_{\mathrm{LIV}}^{(\gamma\textrm{-decay})}=M_{\mathrm{Pl}}/4\lvert\chi\rvert$ then emerges directly from any observed ultra-high-energy cosmic photon~\cite{Konopka:2002tt,Jacobson:zh}. Actually, the observations of multi-TeV events from the Crab Nebula~\cite{Martinez-Huerta:2016azo} have already cast stringent limits on photon decay, with the LIV parameter constrained to the level $\lvert\chi\rvert\lesssim 10^{-2}$, thanks to HEGRA~\cite{Aharonian:2004gb,Schreck:2013paa} and H.E.S.S. experiments~\cite{Aharonian:2006ws,Klinkhamer:2008ky}. The strongest result to date comes from recent observations of $\gamma$-rays above $100~\mathrm{TeV}$ with HAWC. In Ref.~\cite{Albert:2019nnn} the HAWC $2\sigma$ ($95\%$ C.L.) LIV constraint turns out to be $E_{\mathrm{LIV}}^{(\gamma\textrm{-decay})}\gtrsim 2.22\times 10^{22}~\mathrm{GeV}$, over 1800 times the Planck scale ($E_\mathrm{Pl}\approx 10^{19}~\mathrm{GeV}$), in such a way that HAWC observations could exclude the linear-order LIV scale of new physics. However, this is not the case, and we point out that these results do not constrain the linear speed of light modification of the type we considered here (\textit{c.f.},~\labelcref{eq:3.13}). The reason is that photons {\it only} possess {\it subluminal} in-vacuo dispersion~\labelcref{eq:3.5} in the quantum-gravitational D-brane foam medium, as characterized by the quantity $\mathrm{LIV}_{\textrm{D-foam}}>0$, without any superluminal propagation predicted. This implies that photons are \textit{stable} (\textit{i.e.}, \textit{no} photon decay is kinematically allowed), and therefore, using the results based on superluminal dominant reactions, \textit{e.g.}, $\gamma$-decays, one cannot derive the related constraints for the QG coefficients $M_{s}/\varsigma_{\gamma,\mathrm{ani}}^{(\textrm{D-foam})}$ within such a model. There is, again, \textit{no} incompatibility between these strong limits with our main result~\labelcref{eq:4.6} presented above. From another point of view, since the photon decay does not even exist in this framework, it is straightforward to expect that no observation would favor such a process to take place for photons at high energies, and therefore any attempts to search for a hard cutoff compatible with the superluminal part of~\labelcref{eq:4.9} in the observed photon spectra would necessarily result in very strong limits on relevant parameter $\chi$ or related energy scale $E_{\mathrm{LIV}}^{(\gamma\textrm{-decay})}$. In this way, we finally get a consistent description with the $\gamma$-decay limits in the stringy space-time foam scenario.

In summary, given the positive indication of Lorentz violation when analyzing the GRB time delays detected by FGST, we observe that the finding in Refs.~\cite{Xu:2016zxi,Xu:2016zsa,Xu:2018ien} can be described in a stringy model of space-time foam, while being compatible with other astrophysical constraints available to date. We indicate that the strong constraints on LIV imposed from quantum gravitational birefringence effects and superluminal dominant phenomena, \textit{e.g.}, LIV photon decays, can be fully understood in this framework as well. This fact in turn serves as an evidence to support the D-particle string foam interpretation of light speed variation we present here. Worthy to mention that, with the improvement of thirteen orders of magnitude over the past two decades, we could expect to have extra improvements on constraining possible vacuum birefringence, as well as those effects due to superluminal LIV. If more stringent limit can be cast in the future, then this would be a strong support of our D-particle foam interpretation, for reasons explained above. Of course, the validity of the work in Refs.~\cite{Xu:2016zxi,Xu:2016zsa,Xu:2018ien} still needs confirmation through future observations of gamma-rays from the \textit{Fermi} telescope or next generation of satellites~(\textit{e.g.}, the incoming \textit{GrailQuest}~\cite{Burderi:2019cta,Burderi:2021gjx}), and the interpretation of such light speed variance also needs to be continuously subject to further scrutiny.

\section{Closing remarks}
\label{Section5}

In this paper, we present an inspiring way to consistently describe the light speed variation suggested previously from analyses of GRB photons~\cite{Shao:2009bv,Zhang:2014wpb,Xu:2016zxi,Xu:2016zsa,Xu:2018ien,Liu:2018qrg} in the framework of D-particle space-time foam. The fact that this theory coincides well with the observations from \textit{Fermi} data to date is rather intriguing and important, not only because we provide a novel perspective on accounting for the finding in Refs.~\cite{Xu:2016zxi,Xu:2016zsa,Xu:2018ien}, while in agreement with other constraints on LIV, but also because these high-energy astrophysical measurements could further validate hopefully the theoretical details of this type of theories. We conclude that the anomalies of the cosmic photon propagation might have a realistic string-theory origin. More importantly, such an explanation for the light speed variation can also provide hints for a very interesting and long-lasting conjecture that space-time may not be smooth, but ``foamy'' at tiny scales, thus could provide an insightful understanding of the nature of our Universe.

%
%

\section*{Acknowledgements}


We are grateful to Tianjun Li and Hao Li for valuable discussions during the developments of this work. This work is supported by National Natural Science Foundation of China (Grant No.~12075003).

\section*{References}

\end{document}